\documentclass[a4paper,11pt]{article}

\begin{document}

\begin{center}
\bf{Two simple problems in semiclassical dense matter physics} 
\end{center}
\begin{center}
V. \v Celebonovi\'c 
\end{center}
\begin{center}
Inst.of Physics,Pregrevica 118,11080 Zemun-Beograd,
Serbia and Montenegro
\end{center} 
\begin{center}
{\it e-mail}: vladan@phy.bg.ac.yu
\end{center}

\begin{center}
{\bf Abstract}
\end{center}
Physics of dense matter is an extremely rich and complex 
scientific field,
mixing laboratory experiments and astronomical observations
with complex calculations.The aim of this lecture 
is to discuss in some detail two simple but important 
problems which have considerable importance for the developement
of theoretical work in semiclassical dense matter physics.

\begin{center}
{\bf Introduction}
\end{center}

The first historically known attempt to study experimentally a material under high presssure dates from the $XVIII$ century. An english "gentleman"  named Mr.Canton tried to compress water to a pressure of the order of $0.1$ GPa (Block et.al.,1980).To his astonishement,he observed a "transformation" of water into ice.In the vocabulary of modern condensed matter physics this would be called a phase transition.For a recent example of a study of water and ice under high pressure see (Sundberg and Lazor,2004).We can only wonder what would Mr.Canton say today,when 15 different phases of water ice are known.  Systematic laboratory high pressure studes started near the end of the $XIX$ century owing to the efforts of P.W.Bridgman at Harvard University (Bridgman,1964).High pressure experiments have been revolutionized by the discovery of the diamond anvil cell (for a recent review see for example Struzhkin,Hemley and Mao,2004). 
Experimental and theoretical research in high pressure (or,more generally,dense matter physics) is pursued throughout the world. 

Scientific research in high pressure physics has started in Serbia at the beginning of the sixties (Savi\'c,1961).In that paper,which was short and simple but initiated an important series of publications,he pointed out an interesting fact: namely,the mean densities of the planets,as derived from observational data known at the time,could be related to the mean solar density by an extremely simple equation.He did not make any hypothesis about the physical nature of this correlation. 
\newpage

In the next four years Savi\'c in collaboration with Ka\v sanin (Savi\'c and Ka\v sanin,1962/65) developed a theory of the behaviour of materials under high pressure.It was later nick-named the SK theory for short;another synonim is "the semiclassical theory". This last term is justified by the fact that this theory uses only the basic notions of atomic and molecular physics,while it is largely founded on laws of classical physics.

The value of this theory is that it offers the possibility of preforming in a simple way calculations which would be prohibitively complex in rigorous statistical mechanics.Logically,this simplicity very often diminishes the precision of the results. 

The basic physical idea of their theory is that subduing a material to sufficiently high pressure leads ultimately to excitation and ionisation of the atoms and molecules which constitute it. 

At the time when they were starting to develop their theory,the study of the influence of external fields on atoms and molecules was not a theoretically well developed subject.However,Savi\'c and Ka\v sanin used experimental data known at the time to support their idea.

Namely,Bridgmann's experiments were showing that in some materials under certain values of external pressure first order phase transitions occur.It was also known from the analysis of the propagation of seismic waves that their speed changes abruptly a certain depths in the interior of the Earth. Details on their theory are readily avaliable in the literature (for example,\v Celebonovi\'c,1995).

In the remainder of this contribution,we shall discuss at some length two problems of considerable importance for the foundations of the SK theory: 

the relationship between the mean planetary densities derived from observation and the mean solar density,and 

the calculation of the phase transition points in this theory. 

\begin{center}
{\bf The density problem} 
\end{center}
The basic aim of (Savi\'c,1961) was to derive some form of relationship between the solar density and the mean planetary densities.The densities were calculated from the values of planetary masses and radii known at the time.It was shown in that paper that it was possible to fit the values of $\rho$ by the following simple expression:
\begin{equation}
	\rho=\rho_{0} 2^{\phi}
\end{equation} 
where $\rho_{0}=4/3$ is the value of the mean solar density known at the time expressed in CGS units. By suitably assigning to the exponent $\phi$ various {\bf integral} values it turned out to be possible to reproduce known values of the planetary densities.This fit is extremely important for the complete developement of the SK theory because it determines the ratio of densities in two successive phases of a material under high pressure. However,as eq.(1) was derived more than 40 years ago,the logical question about the possibilities of its improvement or even change has started to appear. In the meantime the values of planetary and solar densities have been improved,so this also presented a challenge to eq.(1). 

The following table contains modern values of the mean densities of various objects within the Solar System.All the values given in the table were taken from the internet address http://nssdc.gsfc.nasa.gov/planetary/
planetary home.html . The column denoted by $\rho_{mean}$ contains densities of all the planets of the Solar System and their most important satellites  derived from their observed masses and diameters.The values of the $\phi$ in the third column of table 1 have been obtained by rounding off values calculated from eq.(1).

\begin{table}[t]
\caption{Densities of planets and satellites}
\label{tab:Densities of planets and satellites}
\begin{center}
\begin{tabular}{|c|c|c|c|}
\hline
object & $\rho_{mean}$ & ${\phi}$ & $\rho_{calc}$ \\
\hline
Sun & 1408 & 0 & 1408 \\
Mercury & 5427 & 2 & 5632\\
Venus & 5243 & 2 & 5632\\
Earth & 5515 & 2 & 5632\\
Moon & 3350 & 5/4 & 3349\\
Mars & 3933 & 3/2 & 3982\\
Phobos & 1900 & 2/5 & 1858\\
Deimos & 1750 & 1/3 & 1774\\
Jupiter & 1326 & -1/10 & 1314\\
JI & 3530 & 7/5 & 3716\\
JII & 3010 & 1 & 2816\\
JIII & 1940 & 1/2 & 1991\\
JIV & 1840 & 2/5 & 1858\\
Saturn & 687 & -1 & 704\\
Titan & 1881 & 2/5 & 1858\\
Uranus & 1270 & - 1/7 & 1275\\
Ariel & 1670 & 1/4 & 1674\\
Neptune & 1638 & 1/4 & 1674\\
Triton & 2050 & 1/2& 1991\\
Pluto & 1750 & 1/3 & 1774\\
Charon & 2000 & 1/2 & 1991 \\
\hline
\end{tabular}
\end{center}
\end{table}
   
The column with the heading $\rho_{calc}$ contains values of the density calculated by eq.(1) and the values of $\phi$ given in the table.

Several conclusions can be drawn by simple inspection of Table 1.

The aim of this work has been to test eq.(1) on modern data and on a larger sample of bodies than the one used in (Savi\'c,1961). Eq.(1) proposed by Savi\'c and Ka\v sanin more than 40 years ago has been confirmed on a much larger sample of objects.Note that {\bf non-integral} values of the exponent $\phi$ occur in Table 1. This marks a considerable difference with the original paper by Savi\'c,in which only {\bf integral} values of $\phi$ occured.A possible interpretation of this result can perhaps be found in considerations of the similarities between atomic sturcture and the structure of the planetary system. The physical meaning  of eq.(1) is still a problem open for discussion,but the important concusion of the present work is that the fit proposed by Savi\'c and Ka\v sanin is valid for a much larger set of objects,under the condition that the exponent $\phi$ takes  on non-integral values.

\begin{center}
{\bf The phase transition points}
\end{center}

The theory proposed by Savi\'c and Ka\v sanin predicts that a material subdued to increasing pressure undergoes a sequence of first order phase transitions,numbered by an integer index $i$.The algorithm for the calculation of these values of pressure is presented in detail in previous papers on the subject (for example \v Celebonovi\'c,1995 or later publications).Just for illustration,the final expression for the phase transition pressure is
\begin{equation}
	p_{tr}=0.5101p_{i}^{*};   i = 1,3,5,..
\end{equation} 
and 
\begin{equation}
	p_{tr}=0.6785p_{i}^{*};   i = 2,4,6...
\end{equation}
where
\begin{equation}
p^{*}_{i}\cong 1.8077 \beta_{i} (\overline{V})^{- 4/3} 2^{4 i/3} MBar
\end{equation}
where $\overline{V}$ is the molar volume of the material under study,and the symbol $\beta_{i}$ denotes a constant determined within the theory for each phase of the material.  

However,the problem with eq.(2),and in general with various calculations within this theory is that they have a weak contact with the well established theories of statistical physics. In the remaining part of this paper one such connection will be proposed.

The starting point of all rigorous considerations in statistical physics is the Hamiltonian of the system under consideration.Starting from it,one can derive the expression for the partition function,and from it all the thermodynamic potentials. Accordingly,in attempting to link the theory proposed by Savi\'c and Ka\v sanin with modern statistical physics,one should start from a Hamiltonian (or a suitable expression for one of the thermodynamic potentials) which can in principle predict a number of phase transitions of first order.The order of the transitions is important here because the theory we are discussing here is limited to first order transitions.In the Landau theory of phase transitions,the expression  for the Gibbs potential per unit volume of a system having spontantenous magnetization $M$ has the form (for example,LeBellac,1988):  
\begin{equation}
	g(M)=\frac{1}{2!}r_{0}(T)M^{2}+\frac{1}{4!}u_{0}M^{4}+\frac{1}{6!}v_{0}M^{6}
\end{equation}

where $u_{0}<0$, $v_{0}>0$  and $r_{0}$ has the form $r_{0}(T)=a (T-T_{0})$ ; $T_{0}$ denotes some critical temperature relevant for the problem under consideration and $a$ is a constant. 

It can be shown that $g(M)=0$ for 

\begin{equation}
	M^{2}=-\frac{15u_{0}}{v_{0}}\mp \frac{[900u_{0}^{2}-1440r_{0}v_{0}]^{1/2}}{2v_{0}}
\end{equation}

Clearly,this expression is real if $u_{0}<0$,$v_{0}>0$ and $u_{0}>[(\frac{8}{5})r_{0}v_{0}]^{1/2}$.

It follows from eq.(4) that $\partial g /\partial M =0$ for

\begin{equation}
	M^{2}=-\frac{10 u_{0}}{v_{0}}\mp\frac{[400u_{0}^{2}-480r_{0}v_{0}]^{1/2}}{2v_{0}}
\end{equation}

which implies that $u_{0}^{2}\geq(\frac{6}{5})r_{0}v_{0}$. In a similar way one could determine the position of the points in which the second derivative of $g(M)$ becomes equal to zero.

Replacing the magnetization $M$ in eqs.(3)-(5) by the pressure $P$ renders these expressions applicable to the case of materials under high pressure.In this way a link between the theory proposed by Savi\'c and Ka\v sanin and the Landau theory of phase transitions becomes possible.Eq.(2) contains an experimentally known parameter $(\overline{V})$ and another one derived within the theory $(\beta)$. 

Combining eqs.(2),(3) and (6) and multiplying out all the pure numbers,it follows that for $i=1$: 

\begin{equation}
- \frac{15 u_{0}}{v_{0}}- \frac{[900u_{0}^{2}-1440r_{0}v_{0}]^{1/2}}{2v_{0}}= 2.14258 \beta_{1}^{2}(\overline{V})^{- 8/3}  
\end{equation} 

where value of the constant $\beta_{1}$ is known from the theory of Savi\'c and Ka\v sanin.This means that eq.(8) gives a direct connection between this theory and the Landau theory of phase transitions.Reasoning in the same way for $i=2$,one could obtain a similar equation.Solving a system of these two equations one could obtain the values of the parameters of the Landau theory expressed in terms of the parameters of the theory of Savi\'c and Ka\v sanin.
This calculation is algebraically tedious but straightforward.As a result one would obtain expressions for  $u_{0}$ and $v_{0}$ in terms of $r_{0}(T)$ and the parameters which enter the calculations within the SK theory.  
\newpage
\begin{center}
{\bf Conclusions}
\end{center}

In this lecture two apparently simple problems in dense matter physics have been addressed to some extent. Both of the problems discussed are related to the foundations of the theory of behaviour of materials under high pressure proposed by P.Savi\'c and R.Ka\v sanin nearly fifty years ago. The general conclusion is that one of the basic statements of the SK theory is validated with modern data,and that this theory can be connected with modern theories of phase transitions. 

The simple expression proposed in (Savi\'c,1961) which links the mean planetary densities with the mean solar density was 
tested on modern data.
All the planets and several satellites 
of each of them were included - 21 object in all.
It was shown
that the original fit (eq.(1) of the present paper) is 
confirmed with modern data.However,the exponent $\phi$ in 
eq.(1) is no longer {\bf integral} but can also take 
{\bf non integral} values. This conclusion,although important for the SK theory,opens up the problem of the interpretation of non-integral values of the exponent $\phi$. 

In order to make a contact between the SK theory and some of the better known theories of phase transitions,it was combined with the Landau theory of first order phase transitions.The result is that the Landau theory also has several first order phase transition points,and that the parameters of the Landau theory can be expressed in terms of the parameters of the SK theory.

\begin{center}
References 
\end{center}

Block,S.,Piermarini,G.et Munro,R.G.:1980,{\it La Recherche},{\bf 11},p.806. 

Bridgman,P.W.:1964,{\it Collected Experimental Papers} 

{\bf I-VIII} Harvard Univ.Press.Cambrige Mass.

\v Celebonovi\'c,V.: 1995,{\it Bull.Astron.Belgrade},
{\bf 151},37 

and astro-ph/9603135. 

LeBellac,M.:1988,{\it Des ph\'enom\`enes 

critiques aux champs de jauge},InterEditions/Editions 

du CNRS,Paris.

Savi\'c,P.: 1961,{\it Bull.de la classe des Sci.Math.

et natur.de l'Acad.Serbe des Sci et des Arts},

{\bf XXVI},p.107.

Savi\'c,P. and Ka\v sanin,R.:1962/65,{\it The behaviour of 

materials under high pressure},ed.by Serb.acad.sci.and 

arts,Beograd. 

Struzhkin,V.V.,Hemley,R.J.and Mao,H.K.: 2004,

{\it J.Phys.:Condens.Matter.},
{\bf 16},S1071-S1086. 

Sundberg,S.and Lazor,P.:2004,{\it J.Phys.:Condens.Mater},{\bf 16},

S1223-S1233.

\end{document}